\documentclass[conference]{IEEEtran}

\usepackage[cmex10]{amsmath}
\usepackage{cite}
\usepackage{url}
\usepackage{color}

\usepackage{amssymb}
\usepackage{path}
\usepackage{verbatim}
\usepackage{algorithm}
\usepackage{algpseudocode}
\usepackage{framed}

\usepackage{color}
\usepackage{amsmath}
\usepackage{cases}
\usepackage{subfigure}
\usepackage{theorem}
\usepackage{subfig}
\usepackage{subfloat}
\usepackage{caption}

\usepackage{multirow}
\usepackage{graphicx}
\DeclareMathOperator*{\sometext}{minimize} 
\usepackage{pgfplots}
\usepackage{pgfplotstable}
\pgfplotsset{compat=newest}

{ \theorembodyfont{\normalfont} 

}

\newcounter{enumctr}

\DeclareFontFamily{U}{mathx}{\hyphenchar\font45}
\DeclareFontShape{U}{mathx}{m}{n}{<-> mathx10}{}
\DeclareSymbolFont{mathx}{U}{mathx}{m}{n}
\DeclareMathAccent{\widebar}{0}{mathx}{"73}

\begin{document}

\title{Graph-PHPA: Graph-based Proactive Horizontal Pod Autoscaling for Microservices using LSTM-GNN}

\author{\IEEEauthorblockN{Hoa X. Nguyen, 
		Shaoshu Zhu, and Mingming Liu}
	\IEEEauthorblockA{School of Electronic Engineering, SFI Insight Centre for Data Analytics, Dublin City University, Ireland \\
	mingming.liu@dcu.ie}}
\maketitle

\begin{abstract}
Microservice-based architecture has become prevalent for cloud-native applications. With an increasing number of applications being deployed on cloud platforms every day  leveraging this architecture, more research efforts are required to understand how different strategies can be applied to effectively manage various cloud resources at scale. A large body of research has deployed automatic resource allocation algorithms using reactive and proactive autoscaling policies. However, there is still a gap in the efficiency of current algorithms in capturing the important features of microservices from their architecture and deployment environment, for example, lack of consideration of graphical dependency. To address this challenge, we propose Graph-PHPA, a graph-based proactive horizontal pod autoscaling strategy for allocating cloud resources to microservices leveraging long short-term memory (LSTM) and graph neural network (GNN) based prediction methods. We evaluate the performance of Graph-PHPA using the Bookinfo microservices deployed in a dedicated testing environment with real-time workloads generated based on realistic datasets. We demonstrate the efficacy of Graph-PHPA by comparing it with the rule-based resource allocation scheme in Kubernetes as our baseline. Extensive experiments have been implemented and our results illustrate the superiority of our proposed approach in resource savings over the reactive rule-based baseline algorithm in different testing scenarios.
\end{abstract}

\IEEEpeerreviewmaketitle

\section{Introduction}

Microservices is a new architectural approach that can be applied to cloud-native applications consisting of a collection of small software services. In a microservice architecture, each service has its own functionality, and they jointly contribute to the whole operation of an application deployed on the cloud. Such an application is loosely decoupled into several small services which can be independently deployed on a potentially different platform and technological stack \cite{balalaie2016microservices}. This architectural style allows  cloud resources to be flexibly allocated to each service rather than simply allocating all precious resources to a monolithic application which is often less efficient. More specifically, cloud resources, such as vCPUs, cores and RAMs can be requested and allocated in a scalable and containerised manner both horizontally and vertically providing a flexible control strategy for effective resource management, particularly in a deployment environment subject to some specific service-level-objectives (SLOs), e.g., matching workload with the required resource,. 

Proactive methods, such as machine learning methods, are being widely developed to harmonise resource provisioning for microservice applications. These methods allow different services to be dynamically scaled in a predictive manner, and recently they have demonstrated higher efficiency in scaling and faster response compared to many reactive methods for cluster operations \cite{li2021rambo,sachidananda2021learned,khaleq2021intelligent,gias2019atom}.
More specifically, machine learning algorithms can be applied not only to model the tendency of workloads but also to capture patterns for resource consumption in microservice applications. Furthermore, by leveraging multiple machine learning algorithms, functionalities can complement each other and the benefits can be combined towards a more efficient autoscaling strategy for cloud resources, which has become an emerging research area in recent years. All the above studies so far, however, have not utilized graph-based approaches. Very limited work has been found using graphs to capture important features of microservices from their architecture and deployment environment \cite{park2021graf, qiu2020firm}. 

Motivated by the fact that Graph Convolution Networks (GCNs) have been widely used in other domains and applications, such as transportation and energy \cite{chen2021comparative, liao2021review}, with promising efficacy in describing potential graphical dependency between entities in the network, we aim to leverage a graph-based approach in this paper to design a proactive horizontal pod autoscaling strategy for microservices. With this in mind, the main contributions of our work can be summarized are as follows.
\begin{itemize}
	\item[1. ]  We propose Graph-PHPA, a two-stage prediction method using both Long Short-term Memory (LSTM) and GCN, where LSTM is used for workload prediction and GCN is used to model the relation between workload and resource consumption for different microservices in the network.
	\item[2. ]  We evaluate the performance of Graph-PHPA in a dedicated testing environment composing components from Amazon Web Services  \cite{amazon2022}, Prometheus \cite{prometheus2022} , Grafana \cite{grafana2022} and the Bookinfo application \cite{bookinfo2022}, using real-time workload generated based on a realistic dataset in \cite{zhang2021faster}. 
	\item[3.]   We demonstrate the superiority of Graph-PHPA over the default reactive rule-based method in Kubernetes through extensive experimental studies in different setups.  
\end{itemize}

The rest of this paper is organised as follows. The problem statements and the mechanisms of predictive models are introduced in Section  \ref{problemformulation}, where the proposed solutions are discussed in detail. Experimental setups are introduced in Section \ref{setup} and results are discussed in Section \ref{results}. Finally, the conclusion for the current research and potential future research directions are outlined in Section \ref{conclusion}.

\section{Problem formulation and Models} \label{problemformulation}

\subsection{Problem statement}

We consider a scenario where $n$ services are deployed in a Kubernetes cluster with constrained resources for vCPU share. The vector, $\mathbf{N}{(t)} := [N^{(t)}_1, N^{(t)}_2, ..., N^{(t)}_n]$, denotes the number of pods for each of the microservice at time $t$. Note that $ 1 \leq N^{(t)}_i \leq Q_i, \forall i = 1,...,n$, where $Q_i$ is the upper bound of pods that the microservice $i$ can have. The vCPU share vector, $\mathbf{R}{(t)} := [R^{(t)}_1, R^{(t)}_2, ..., R^{(t)}_n]$, represents the vCPU for each microservice at time $t$. Note $R_i^{lb} \leq R_i^{(t)} \leq R_i^{ub}, \forall i = 1,...,n$, where $R_i^{ub}$ and $R_i^{lb}$ denote the upper and lower vCPU bounds.

Our key objective is to find out the number of pods $\mathbf{N}{(t)}$ for microservices based on their vCPU shares $\mathbf{R}{(t)}$ considering the dependence among microservices and the workload input. Once the workload changes, the system can dynamically add resources for the microservices to ensure the application's performance or reduce unnecessary replicas to save operational cost. In the following sections, we shall introduce the design of predictive models in detail, which consists of the predictive model for workload, the predictive model for resource usage, and the integrated model, i.e., the proposed Graph-PHPA. 


\subsection{Predictive models}

\paragraph{Predictive model for Workload}

Let $a_i^{(t)}$ denote the workload of microservice $i$ at time $t$. Let $ \mathbf{a}_i^{(t)} := [a_i^{(t-k+1)}, ..., a_i^{(t-1)}, a_i^{(t)} ]$ denote the workload feature vector of the microservice $i$ at time $t$ for a given time window with size $k$, where $t \in [k, T]$ and $T$ denotes the length of input data. Given the vector $\mathbf{a}_i^{(t)}$, we aim to forecast the workload $\widetilde{a^{(t+1)}_i}$ of microservice $i$ at time $(t + 1)$. The LSTM model \cite{gers2002applying} is applied here to address the time series prediction problem. More specifically, we construct the LSTM model consisting of multi-layers. The cells of each layer are connected with each subsequent one. The input layer receives the sliced data for the corresponding time window. The last layer is a dense layer using the ``tanh'' activation function to produce the predicted workload value. Mathematically, we wish to find a learning function $\psi(.): \mathbb{R}^{k} \mapsto \mathbb{R}$ which is able to address the following optimization problem through a multi-layer LSTM network: 
\begin{equation}\label{eq:1}
\begin{split}
    \sometext_{\psi} & ~ \sum_{t=k}^{T-1} (a^{(t+1)}_i - \widetilde{a_i^{(t+1)}})^2 \\
    s.t. \quad \psi(\mathbf{a}_i^{(t)}) & = \widetilde{a_i^{(t+1)}}.
\end{split}
\end{equation}

\paragraph{Predictive Model for Resource usage}
We leverage Graph Convolution Network (GCN) to accommodate the graphical structure of the application and the dependency of microservices. The graph structure of microservices for testing application is assumed to be static and deterministic throughout our experiments. The goal of the model is to learn a function of features on a graph $\mathcal{G} = (\mathcal{V},\mathcal{E})$ which takes features of each node $x_i: x \in X^{N \times D} $ and a representative description of the graph structure in adjacency matrix $A \in \mathrm{R}^{N \times N}$, where $X$ is the feature matrix, $N$ is the number of nodes $v_i \in \mathcal{V}$, edges between nodes $(v_i, v_j) \in \mathcal{E}$,  $D$ is the number of input features. The output $Z^{N \times F}$ is a node-level feature, where $F$ is the number of output features. In each layer of $L$ number of layers, the output of the layer is $H^{(l+1)} = \sigma(H^l, A)$, where $H^{(0)} = X$ and $H^{(L)} = Z$. The chosen activation function $\sigma(.,.)$ and the parameters decide the differences among the GCN models. We use the layer-wise propagation rule introduced in \cite{kipf2016semi, bloemheuvel2022multivariate} as follow:
\begin{equation}\label{eq:2}
    H^{(l+1)} = \sigma \left( \Tilde{D}^{-1/2} \Tilde{A} \Tilde{D}^{-1/2} H^{(l)} W^{(l)} \right),
\end{equation}
where, $\Tilde{A} = A + I_N$ is the adjacency matrix of the graph $\mathcal{G}$. $I_N$ which is the identity matrix representing self connections of each node. $\Tilde{D_{ii}} = \sum_j \Tilde{A}_{ij}$ is the diagonal node degree matrix and $W^{(l)}$ is a layer specific trainable weight matrix. $\sigma(.)$ represents an activation function, for example the $ReLU(.) = max(0,.)$. $H^{(l)} \in \mathrm{R}^{N \times D}$ is the matrix of activation in the $l^{th}$ layer, $H^{(0)} = X$ in the first layer.   
Let $b_i^{(t)} \in \mathbb{R}$ be the maximum resource consumption of microservice $i$ at time $t$ over a time window of size $k$. In other words, $b_i^{(t)}$ represents the maximum amount of resource required for microservice $i$ given the workload $\mathbf{a}_i^{(t)}$. Our objective is to find a learning model $\phi(.): \mathbb{R}^{k} \mapsto \mathbb{R}$ which can address the following optimization problem:
\begin{equation}\label{eq:3}
\begin{split}
    \sometext_{\phi} ~ \sum_{t=k}^{T-1} (b^{(t+1)}_i - \widetilde{b_i^{(t+1)}})^2 \\
    s.t. \quad \phi([a_i^{(t-k+2)}, ... ,a_i^{(t)}, \widetilde{a_i^{(t+1)}}]) & = \widetilde{b_i^{(t+1)}}.
\end{split}
\end{equation}

\noindent where, $\widetilde{b_i^{(t+1)}}$ is the predicted maximum amount of resource for microservice $i$ at $t+1$ over a time window of size $k$. 
The prediction model takes the historical workload [$a_i^{(t-k+2)}, \ldots, a_i^{(t-1)}$], current workload $a_i^{(t)}$ and the predicted workload $\widetilde{a_i^{(t+1)}}$ from \eqref{eq:1} as input and outputs the desired amount of resource, i.e., vCPU share, for the given microservice $i$ for the next time step. 

\paragraph{Integration algorithm}
Algorithm \ref{alg:1} presents the proposed Graph-PHPA algorithm, which assembles the two predictive models. The algorithm works as follows. At each time step $t$, \( \mathbf{a}^{(t)}_i\) is collected from the monitoring system and is used to predict \( \widetilde{a_i^{(t+1)}} \). Then, the model $\phi$ predicts \( \widetilde{b_i^{(t+1)}} \) using $[a_i^{(t-k+2)}, ... ,a_i^{(t)}, \widetilde{a_i^{(t+1)}}]$. By that, resource prediction model accounts for the future workload. For example, current resource of a microservice is $R_i^{(t)} =$ 2.0 vCPU at $t$. However, the predicted resource \( \widetilde{b_i^{(t+1)}} \) requires 2.5 vCPU, which implies that an extra 0.5 vCPU is required to be prepared for the service $i$ at $t$ for the upcoming workload at time $(t+1)$ as per the prediction. Thus, 0.5 vCPU, e.g., 1 pod (assuming 0.5 is the unit specification), is to be provisioned at time $t$ to avoid delay in the initialization process of pods in practical operations. Therefore, the number of pod $N_i^{(t+1)}$ is updated accordingly based on the predicted resource $\widetilde{b_i^{(t+1)}}$ as per the resource specification for each microservice $i$ denoted by $v_i^{p}$. 
\begin{algorithm}
\caption{Integration Algorithm - Graph-PHPA} \label{alg:1}
\begin{algorithmic}
\State \textbf{Input:} Window size for forecasting $(k)$, learned workload prediction model $\psi: \mathbf{a}^{(t)}_i \rightarrow \widetilde{a_i^{(t+1)}}$, learned resource prediction model $\phi:[a_i^{(t-k+2)}, ..., a_i^{(t)},\widetilde{a_i^{(t+1)}}] \rightarrow \widetilde{b_i^{(t+1)}}$, allocated vCPU resource to microservice $R^{(t)}_i$, standard vCPU resource specification of microservice $i$ for one pod $v_i^p$, allocated number of pods for the microservice $i$, $N^{(t)}_i$. \\
\State \textbf{Output:} Predicted vCPU shares $R_i^{(t+1)}$ and number of pods $N_i^{(t+1)}$ for each microservice $i$ at time $t+1$. \\
\For{$t = 1,2,3,...$}
\State $\widetilde{a_i^{(t+1)}} \gets $  $\psi(\mathbf{a}^{(t)}_i)$
\State $\widetilde{b_i^{(t+1)}} \gets $  $\phi([a_i^{(t-k+2)}, ..., a_i^{(t)},\widetilde{a_i^{(t+1)}}])$
\State $ R_i^{(t+1)} \gets \widetilde{b_i^{(t+1)}}, \textrm{where} ~ R_i^{lb} \leq R_i^{(t+1)} \leq R_i^{ub}$
\If{$R_i^{(t)}  < R_i^{(t+1)}$ }
    \State $N_i^{(t+1)} \gets N_i^{(t)} + \bigg \lceil \frac{R_i^{(t+1)} - R_i^{(t)}}{v_i^p} \bigg \rceil, \forall i = 1, 2, \ldots, n $  \\
    \State $N_i^{(t+1)} \gets \textrm{min}(N_i^{(t+1)}, Q_i)$ \\
    \Comment{Adding some pods}
\ElsIf{$R_i^{(t)} > R_i^{(t+1)}$ }
    \State $N_i^{(t+1)} \gets N_i^{(t)} - \bigg \lceil \frac{R_i^{(t)} - R_i^{(t+1)}}{v_i^p} \bigg \rceil, \forall i = 1, 2, \ldots, n $ \\
    \State $N_i^{(t+1)} \gets \textrm{min}(N_i^{(t+1)}, Q_i)$ \\
     \Comment{Reducing some pods}
\EndIf
\EndFor
\end{algorithmic}
\end{algorithm}

\section{Experimental setup} \label{setup}
\subsection{Microservice Application Deployment}

In this section, we first present our dedicated environment for deployment and testing of the microservice application, and then we discuss load generation, baseline algorithm, and evaluation methods. We evaluate our algorithm on an interactive and practical real-world microservice application, namely Bookinfo \cite{bookinfo2022}. Bookinfo is a representative microservice application in the research community. The structure of the application consists of four microservices implemented using different technological stacks. The details of the microservices are outlined as follows: 

\begin{itemize}
	\item \textit{productpage} microservice transfers requests to the \textit{details} and \textit{reviews}.
	\item \textit{details} microservice stores the book information.
	\item \textit{reviews} has three versions which contain reviews of the book. Both version v2 and version v3 invoke the \textit{rating} and return black and red stars, respectively. Version v1, however, doesn't call \textit{ratings}.
	\item \textit{ratings} has the ranking info of books.
\end{itemize}

Bookinfo is a polyglot application, e.g., $reviews$ is written in Java, while $details$ is written in Ruby. Our experimental environment consists of a test server and the system under test, i.e., the Bookinfo application. The experimental application is deployed on Amazon Web Services (AWS) \cite{amazon2022}. We used c5.xlarge EC2 instances for the test server with numerical resource limitation of  4 cores vCPU, 3.4 GHz, 8G memory, 100G storage. We used Kubernetes \cite{kubernetes2022} as the container orchestration engine for four microservices subjected to various workload scenarios generated by the open-source load testing tool named Locust \cite{locust2022}. The Kubernetes manager can host 79 pods at maximum, where each node equally hosts the pods in the cluster thanks to the load balancer component. 

\begin{figure}[htbp]
\centerline{\includegraphics[width=8cm]{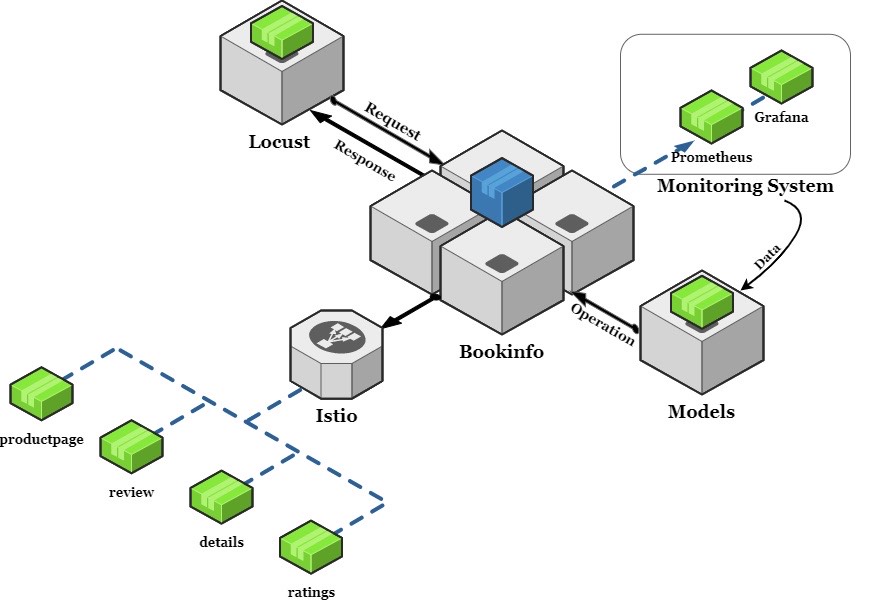}}
\caption{Experimental environment.}
\label{fig:13}
\end{figure}

Common Python libraries, including Pytorch \cite{paszke2019pytorch}, Tensorflow \cite{abadi2016tensorflow}, and Keras \cite{chollet} were used to build our prediction models. The Graph-PHPA loads the trained models and processes performance metrics to make a prediction on the desired number of pods for running in the next step, and this information is sent to the Kubernetes scheduler for scheduling. Fig. \ref{fig:13} presents a schematic diagram for our experimental environment. The test server was deployed on AWS EC2 and the system under test (SUT) was deployed on AWS EKS. The Locust server was hosted by Docker and all the microservices in Bookinfo were injected by Envoy Sidecar service controlled by Istio for monitoring and load balancing. We then used Prometheus \cite{prometheus2022} and Grafana \cite{grafana2022} to monitor and collect the testing data (resource usages and workload data) for each microservice. The monitoring data was collected every minute consisting of vCPU, memory, and workload for each microservice at one minute interval. 

\subsection{Load generation}
To evaluate the Graph-PHPA, we used a practical workload trace in our experimental environment, which is a trace of function invocations in Microsoft's Azure Functions \cite{zhang2021faster} from 31/01/2021 to 13/02/2021 with time interval of 5 minutes. We  extracted and scaled the invocation calls in the dataset to mock the requests to cloud cluster. The workload trace is available publicly and details can be found in \cite{park2021graf}. Locust consequently spawned the number of requests every minute to stress microservices in Bookinfo application. 

\subsection{Evaluation of the prediction model}
We evaluate Graph-PHPA in two steps. First, we train two separate models using LSTM and GCN based on the experiment data collected from Granfana. We assess the workload prediction model and the resource prediction model using MSE/MAE to identify the best hyperparameters for the models. The next phase is to predict the incoming workload and consequently the required vCPU of the next step using the above trained models. The Graph-PHPA is evaluated on real-time workload \cite{zhang2021faster}. The results are then compared with the baseline algorithm using Kubernetes HPA. The evaluation results of each step are discussed in the following sections. 

\paragraph{Training predictive models}
The training data is collected from our testbed to train the predictive models. The models are trained with various settings and the model with the minimum MSE is used for Graph-PHPA. The whole dataset is divided into a training set (60\%), validation set (20\%), and test set (20\%), i.e., the last 800 data points to be used in our result analysis. We used the workload prediction model to predict the number of requests for the microservices. At $t$, the LSTM model predicted the workload for the next time step $t+1$ based on the last $k$ observations. We investigated various window sizes $k$, including 4, 6, 8, and 10 to identify the best input window size for the LSTM model. We chose the input window size of 10 for the best performance. The prediction algorithm was trained with the following hyperparameters, including optimizers (Adam at learning rate (lr) = 0.01), epochs (50), batches (64), number of units (10, 50, 75, 100, 125, 150, 175, 200), and number of hidden layers (1, 2, 3, 4, 5). The model performance was assessed and compared using MSE and MAE on the test dataset. 

The settings of the resource prediction model include the input features, neural network structure, and loss function. We carefully chose these parameters to obtain an accurate prediction model. The node feature was the workload of each microservice. The workload data over a fixed-length window $k$ was then fed into the resource prediction model as input for predicting the maximum resource consumed across the window. The adjacent matrix is computed with the assumption that the graph structure of the application is undirected with self-loops. The setup which yields the minimum MSE on the test dataset was chosen as our setup for the Graph-PHPA.  

\paragraph{Graph-PHPA}
We evaluate the Graph-PHPA using the real-time workload generated based on the realistic dataset on our testbed. Each pod has a limit of 1 vCPU ($v_i^p$ = 1 vCPU) and 2GB of memory. Once the workload changes, the Graph-PHPA proactively recalculates the number of pods of each microservice using the trained predictive models, and redeploys these pods to the cluster accordingly. 

\section{Experimental results} \label{results}

\subsection{Workload Forecasting Evaluation Results}

\begin{figure}[htbp]
	\centerline{\includegraphics[width=8cm]{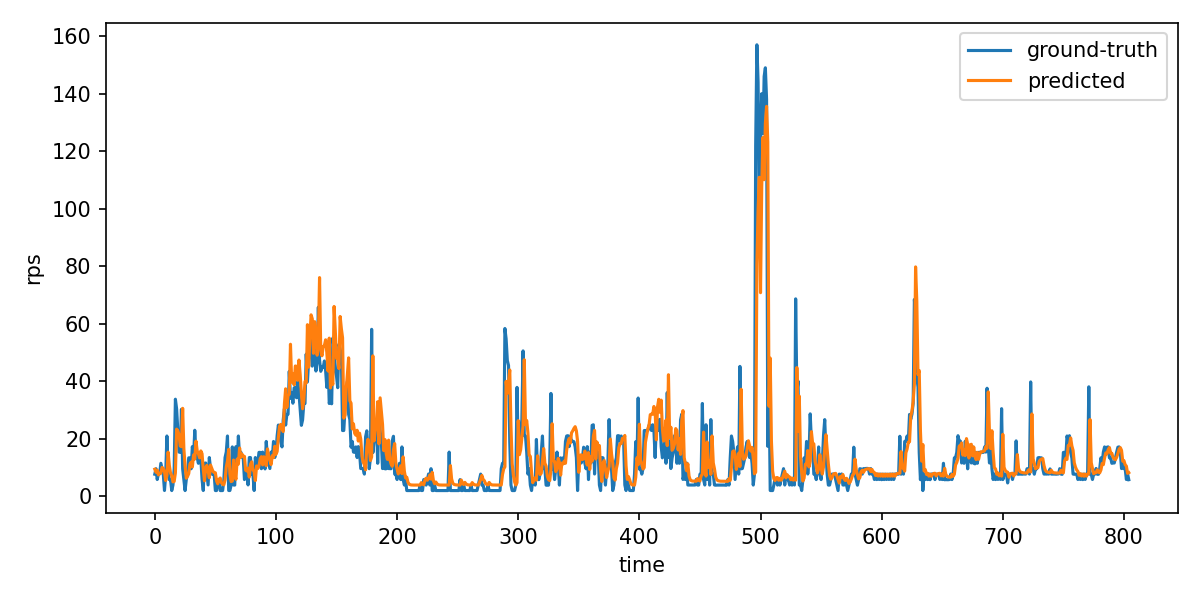}}
	\caption{Workload prediction model.}
	\label{fig:6}
\end{figure}

Fig. \ref{fig:6} shows the ground truth (requests per second) and the predicted workload traces leveraging the best model we found in the work. The best model adopted an Adam optimizer (learning rate equals 0.01) with one LSTM layer of 50 hidden units and a batch size of 64. We can observe that the predict workload value can easily match the ground truth.

\subsection{Resource prediction model}

\begin{figure}[htbp]
	\centerline{\includegraphics[width=0.5\textwidth]{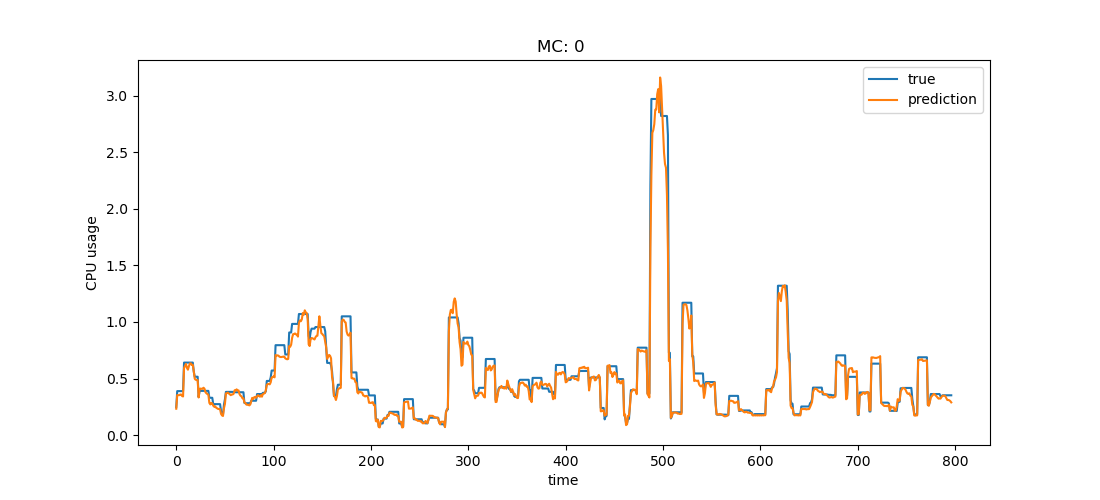}}
	\caption{vCPU prediction for $product page$ microservice.}
	\label{fig:7}
\end{figure}

Fig. \ref{fig:7} illustrates the best resource prediction results for the ``product page'' microservice only as it consumes the highest amount of vCPU resource in the application. Our results have shown that the vCPU usage has been captured quite well for the microservice after fine-tuning the resource prediction model through its hyperparameters: optimizers (Adam at lr = 0.001), epochs (100), batches (256), and two GCN layers. 

\subsection{Overall System Evaluation Results}
\begin{figure}[htbp]
	\centerline{\includegraphics[width=0.6\textwidth, height=2.8in]{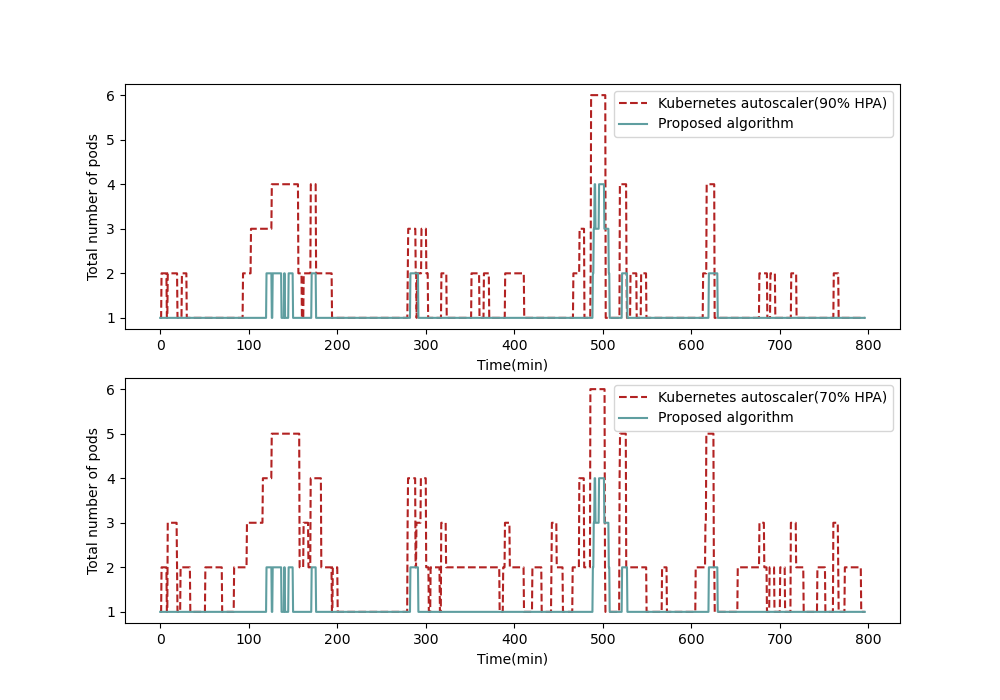}}
	\caption{Total number of pods allocated to $product page$ microservice with respect to vCPU threshold 90\% (upper subplot) and 70\% (lower subplot).}
	\label{fig:50}
\end{figure}

We demonstrate the comparison of our proposed Graph-PHPA algorithm with the baseline algorithm in Kubernetes for the Bookinfo application in Fig. \ref{fig:50}, where only vCPU allocation of product-page microservice is shown. Once the vCPU utilization of any pods is greater than the user-defined value (vCPU threshold), Kubernetes dynamically adds pods to the microservice. On the other hand, Kubernetes removes a pod as the CPU utilization of all pods is less than the scale-in threshold. Our proposed algorithm effectively allocates the required pods to the microservice using the predicted workload and the trained resource allocation model. It is clearly seen in Fig. \ref{fig:50} that the number of pods of $product page$ microservice is adjusted appropriately using the forecasted workload, which shows promising performance of our method compared to the baseline algorithm in Kubernetes under the same real-time test workload profile. More specifically, the lower subplot compares our Graph-PHPA result with that of the Kubernetes autoscaler with 70\% threshold. Compared to the upper subplot using 90\% threshold, the lower subplot shows significant resource saving using the proposed Graph-PHPA method. Although a larger threshold can indeed reduce the number of pods, it also poses the higher risk of overloading especially in a reactively controlled manner, which may be less desirable in any practical operation.  

\section{Conclusion}\label{conclusion}

In this paper, we have presented a novel graph-based proactive horizontal pod autoscaling strategy for microservices using LSTM-GNN, namely Graph-PHPA. This approach is two-stage as it first predicts the upcoming workload by using LSTM and takes this output to further infer the desired amount of resources, i.e., pods. This two-stage architecture allows each individual model to be replaced in a plug-and-play manner for further development. The superiority of our proposed approach has been validated through extensive experiments implemented using a dedicated testing environment with real-time workload. Significant resource savings have been shown compared to the baseline algorithm implemented using the Kubernetes HPA. Finally, we note that one limitation of the current work is that the proposed model only focuses on the prediction of vCPU usage whilst taking account of the workload as the model attribute, which will be addressed as part of our future work. In addition to this, we will also further investigate the efficacy of our proposed algorithm by evaluating it against a group of related algorithms in our future work.

\section*{Acknowledgement}

This work is supported by the Huawei Ireland Research Centre for the scalability and provisioning surveillance project and Science Foundation Ireland (SFI) under Grant Number SFI/12/RC/2289\_P2 (Insight SFI Research Centre for Data Analytics), co-funded by the European Regional Development Fund in collaboration with the SFI Insight Centre for Data Analytics at Dublin City University.

\bibliographystyle{ieeetran}
\bibliography{References}

\end{document}